\documentclass{WileyMSP-template}
\usepackage{amsmath}

\makeatletter
\newenvironment{sqcases}{%
  \matrix@check\sqcases\env@sqcases
}{%
  \endarray\right.%
}
\def\env@sqcases{%
  \let\@ifnextchar\new@ifnextchar
  \left\lbrack
  \def\arraystretch{1.2}%
  \array{@{}l@{\quad}l@{}}%
}
\makeatother

\begin{document}
\sloppy

\pagestyle{fancy}
\rhead{\includegraphics[width=2.5cm]{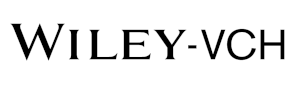}}

\title{Low and high frequency noise in LEDs}

\maketitle


\author{Danylo Bohomolov*}
\author{Vita Ivanova}
\author{Ulrich T. Schwarz}


\dedication{}

\begin{affiliations}
Danylo Bohomolov\\
Chemnitz University of Technology, Str. der Nationen 62, 09111 Chemnitz, Germany\\
National Technical University of Ukraine ``Igor Sikorsky Kyiv Polytechnic Institute'', Beresteiskyi Ave 37, 03056 Kyiv, Ukraine\\
Email Address: danylo.bohomolov@physik.tu-chemnitz.de

Prof. Dr. Vita Ivanova\\
National Technical University of Ukraine ``Igor Sikorsky Kyiv Polytechnic Institute'', Beresteiskyi Ave 37, 03056 Kyiv, Ukraine\\

Prof. Dr. Ulrich Theodor Schwarz\\
Chemnitz University of Technology, Str. der Nationen 62, 09111 Chemnitz, Germany\\
\end{affiliations}



\begin{abstract}

LED degradation is usually associated with defects in the active region. Whereby the noise analysis can be a strong instrument to reveal them. The results of optical noise measurements for commercially available blue LED samples in a wide frequency range from kHz to MHz are reported. Noise spectra were decomposed into components according to the presented theoretical model which includes $1/f$-type noise, generation-recombination noise, and white noise. The $1/f^\gamma$-type noise was modeled as a superposition of generation-recombination noise components at defects with a continuous wide distribution of relaxation lifetimes. The coincidence of the experimental results with $1/f^\gamma$ model for the low frequency range is proved and a fitting is made. We identified three noise components that are highly current-dependent. The corresponding model of temperature dependence for the low-frequency range was developed. At low currents, the model partially matched the experimental results in the temperature range from 100~K to 300~K at low frequencies. However, high-frequency measurements showed deviations from the expected Lorentzian behavior.

\end{abstract}


\section{Introduction}

For the proper operation of almost all light-emitting diode (LED) based devices, the stability and longevity of the light source are desirable. The presence of defects in a crystal reduces the efficiency of a LED and accelerates its aging \cite{koike2002}. Therefore, knowledge of the types and distribution of defects in the structure is extremely important. 

Among the existing methods of defect characterization, e.g. deep level transient spectroscopy or 
scanning transmission electron microscopy, noise analysis is the cheapest and nondestructive. Defects in the LED's active region modulate randomly the free charge carrier number and, as a consequence, lead to the photon number (optical) fluctuations in the same phase as electrical ones~\cite{glemza2018}.
Since the crystal quality of a quantum well is associated with noise at medium currents, in this work we did not focus on experiments at very high currents, where fluctuations of hot charge carriers must be taken into account~\cite{bychikhin2005}.

Noise in the low-frequency region is considered to be the dominant noise in LEDs~\cite{dobrzanski2004}. Its study is a highly sensitive method for assessing degradation, as it is an effective marker of the defect structure of an electronic device \cite{palenskis2021}. The increase in low frequency noise components explicitly correlates with defect formation~\cite{palenskis2021}. Thus, low-frequency noise can act as an indicator of LED reliability~\cite{pralgauskaite2015}. Although the noise in the high-frequency regions is much lower, it is a valuable additional source of information for assessing the defectiveness of the LED structure. 

Our work aims to examine whether the proposed model in general and its components, including the proposed distribution of trap relaxation time constants in particular, are in good agreement with the experimentally obtained noise of LEDs.

\subsection{Noise types in LED}

Noise can be represented as random deviations of a signal from its average value. For LEDs, the point of interest is the fluctuation of output photon flux. This noise corresponds to the fluctuation of free charge carriers in the active region, which are involved in recombination current~\cite{rumyantsev2005,palenskis2011,sawyer2006}. Photon fluctuations imply optical noise.

For quantitative noise evaluation in this work, the power spectral density (PSD) will be used, which describes the distribution of power across the frequency components that constitute the noise signal.

It is well known that any LED is characterized by three main types of noise: white, generation-recombination and $1/f$-type noise~\cite{palenskis2021}.

\subsubsection{White noise}
This is a fundamental type of noise that describes thermal fluctuations of charge carriers (Johnson-Nyquist noise), as well as fluctuations due to the discrete nature of electrons and/or photons (shot noise)~\cite{vanderziel1979}. The sum of this two components is frequency-independent. It has been shown that in the low frequency region white noise is much smaller than frequency dependent noise components, i.e. $1/f$-type and GR noise~\cite{kogan1996}.

\subsubsection{Generation-recombination noise}

Generation-recombination (GR) noise in LEDs is caused by capturing and releasing of free charge carriers. The largest component of this noise is trap-assisted GR noise, discussed in detail in Appendix~\ref{appendix_gr}. For the transition of one type of carrier between single trap and only one band, the PSD is described by \textbf{Equation~(\ref{eq:grdot})}, where $\langle\delta N^2\rangle$ is related to the variance of the number of charge carriers in a sample.

\begin{equation}\label{eq:grdot}
	S_{\text{GR}}(f)=\frac{4\langle\delta N^2\rangle\tau}{1+\tau^2(2\pi f)^2}
\end{equation}

A power spectrum with the form of Equation~(\ref{eq:grdot}) is called a Lorentzian spectrum and describes noise with a characteristic relaxation lifetime $\tau$. This noise has a white shape below the characteristic frequency $1/\tau$ (we will call this part as plateau) but falls off as $f^{-2}$ above this frequency. 

The time constant $\tau$ is the inverse of the sum of emission and capture time constants for a single trap~\cite{machlup1954,vandamme2008,rimini1992,lukyanchikova1997}. These times correspond to the inverse of the derivatives of generation $\text{d}g/\text{d}N$ and recombination rates $\text{d}r/\text{d}N$, respectively~\cite{mitin2002}. 
It has been also shown~\cite{lukyanchikova1997} that in the low frequency region, the relaxation lifetime for deep trap levels coincides with the Shockley-Read-Hall (SRH) time constant (Appendix~\ref{appendix_gr}).

\subsubsection{$1/f$-type noise}

This type of noise, sometimes also called flicker noise, is found in almost all physical systems and has no general explanation~\cite{vanderziel1979}. It is described by a PSD that is proportional to $1/f^\gamma$, where $\gamma\in[1,2)$. It is observed that in LEDs, the amplitude and power $\gamma$ of this noise strongly depend on the presence of defects~\cite{vanderziel1979}. This type of noise is most pronounced in the low-frequency region due to its shape. 
It is known that the drift, and the diffusion, of the charge carriers does not yield $1/f$ noise~\cite{kogan1996}.

\subsection{Model}~\label{subsection_model}

In real materials, the energy levels of the traps are distributed in a certain way, forming a corresponding distribution of energies, and carrier relaxation times.
If there are several trap levels, the noise spectrum is characterized by several relaxation lifetime constants $\tau$. These terms show up as several distinct Lorentzian bumps in the PSD curve. However, if the relaxation times are close to each other, the GR noise spectrum should be obtained by integrating over $\tau$ distributed in a certain way within a given range. We investigated different approaches to representing the probability density function, which significantly affect the shape of the resulting spectrum.

The long-standing debate on the origin of $1/f$-type noise was reflected in the theory of Du Pre and Aldert Van der Ziel~\cite{vanderziel1988} and its physical interpretations, e.g. in McWhorter model, which takes into account the simultaneous action of multiple-trap levels~\cite{mcwhorter1957}. 

In other approaches, flicker noise is modeled by a superposition of processes as random telegraph signal (RTS). There is also a two-level system model to describe the behavior of deep-level point defects that have several metastable states~\cite{kirtonuren1989}. What all existing theories have in common is a purely mathematical procedure for representing flicker noise as a superposition of Lorentzian noise components with certain carriers characteristic relaxation lifetime distribution. Both GR noise and a RTS have the Lorentzian form of PSD. 

In the McWhorter~\cite{mcwhorter1957} model, the required distribution function of the relaxation lifetimes $g(\tau)$ is introduced as $g(\tau)\propto1/\tau$ as the result of carrier tunneling to trap levels within the surface oxide at different depths (depth distribution). This is a successful approximation for field-effect transistors or tunnel diodes devices, for which this model was developed. Such a distribution leads to a $1/f^{\gamma=1}=1/f$ component. Mathematically, other forms of $g(\tau)$ can give other forms of low-frequency $1/f$-type PSD, but such distributions must be physically relevant.

However, this distribution is only partially applicable to LEDs, because it does not explicitly relate to the GR processes in the active region. A more suitable model is an uniform distribution of characteristic relaxation lifetimes.  
Since the density of the vacant traps (for electrons or holes) obeys the Boltzmann distribution~\cite{bisquert2008} and the rates of capture-release processes follow Arrhenius' law, this naturally gives an uniform distribution for the relaxation times, which follows from the conservation of probability $g(E)\,\text{d}E=g(\tau)\,\text{d}\tau$.
Integrating Equation~(\ref{eq:grdot}) and taking into account uniform distribution $g(\tau)\propto1/(\tau_\text{max}-\tau_\text{min})$, we obtain the integral in form of \textbf{Equation~(\ref{eq:gr})}. Detailed derivation is given in Appendix~\ref{appendix_dist}.

\begin{equation}\label{eq:gr}
	S_\text{GR}(f)=A_\text{GR}\int\limits_{\tau_\text{min}}^{\tau_\text{max}}\frac{\tau g(\tau)d\tau}{1+(2\pi f\tau)^2}=
    \begin{sqcases} \text{for\ }\tau_\text{max}\gg\tau_\text{min}: &  
        \begin{cases}
        \text{const} &\text{for}\ f\ll\tau_\text{max}^{-1}\\
        \propto 1/f^{\gamma(f)} &\text{for}\ \tau_\text{max}^{-1}< f<\tau_\text{min}^{-1}\\
        \propto 1/f^{2} &\text{for}\ f\gg\tau_\text{min}^{-1}\\
        \end{cases} \\ 
        \begin{aligned}
        \text{for\ } \tau_\text{max}=\tau_\text{min}(1+\epsilon)\\ \epsilon\to0
        \end{aligned}:
        &\propto\dfrac{\tau}{1+(2\pi f \tau)^2} 
    \end{sqcases}
\end{equation}

%

If the integration limits are wide distributed $\tau_\text{min}\ll\tau_\text{max}$, the integral gives us the flicker noise component proportional to $1/f^{\gamma(f)}$ with gradually $\gamma$ change. In a case of narrow $\tau$ distribution, we get the discrete Lorentzian with an average characteristic relaxation lifetime $\tau$ within a range.

\begin{figure}[h!]
\centering
  \includegraphics[width=0.75\linewidth]{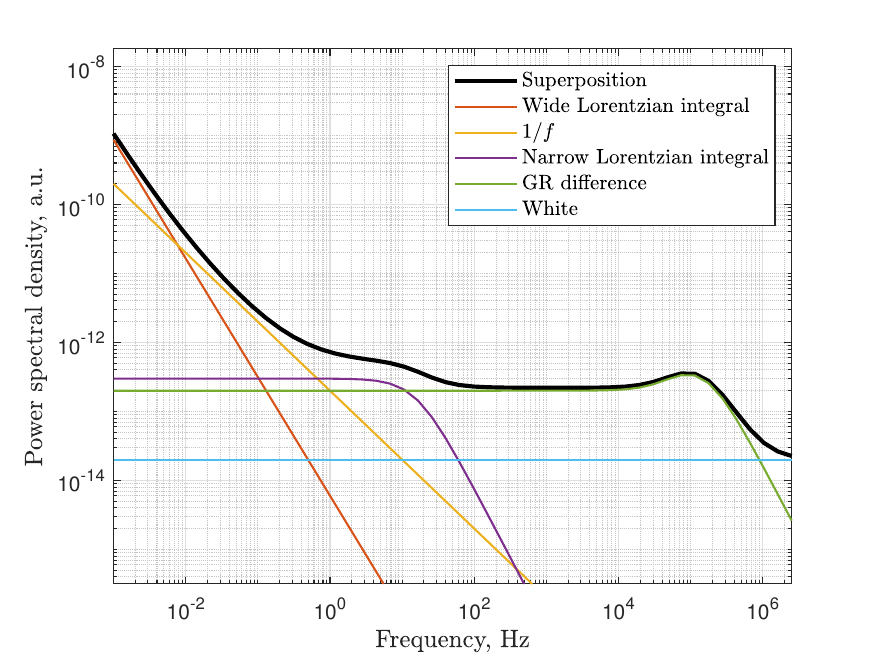}
  \caption{Superposition (black) of three types of components in loglog scale: $1/f$-type noise, GR noise and white noise. The different types include the following components: $1/f$ noise (yellow), {integral} superposition of GR noise components with wide distribution, i.e. $1/f^\gamma$ noise (orange), {integral of} single GR noise component {with narrow distribution (purple)}, sum of two Lorentzians with positive and negative amplitude (green) and white noise (blue), which is the sum of thermal and shot noise.}
  \label{fig:model}
\end{figure}

Since there can be several integrals with different widths and distribution functions, we can observe several Lorentzians and several, typically two with $\gamma=1$ and $1/f^\gamma$ components simultaneously. $1/f$ component is typically observed at low currents and not always present. For fitting simplification instead of taking integral, we will use empirical $1/f^\gamma$ with $\gamma$ limitation. The physical reasons of an additional $1/f$ component 
are beyond the discussion in the article. In such a case, the total PSD model will be in the form of \textbf{Equation~(\ref{eq:model})}, where parameter $A$ represents amplitudes of the corresponding noise components.

\begin{equation}\label{eq:model}
	S_\text{tot}(f)=\frac{A_{1/f}}{f}+\frac{A_{1/f^\gamma}}{f^\gamma}+\sum\limits_i\frac{\tau_i A_{{GR}_i}}{1+(2\pi f \tau_i)^2}+S_\text{white}+S_\text{syst}
\end{equation}
Moreover, a correlation between e and h GR processes is present (what can occur if defect energy levels near the middle of the bandgap), this leads to the Lorentzian with negative amplitude in the sum in Equation~(\ref{eq:model}) appearance~\cite{rimini1992}. In logarithmic scale {Equation~(\ref{eq:model})} is shown in \textbf{Figure~\ref{fig:model}}.

\section{Low-frequency noise of blue LED}

For proof of concept, the test sample is a typical commercially available blue LED with a wavelength of 455 nm. The LED is powered by a constant, fully analog ultra-low-noise current source ranging from 0.5~mA to 200~mA. The optical signal is read at a distance by a Si photodiode with 100 mm$^2$ active area. 

Often, due to setup limitations, it is impossible to simultaneously measure both low-frequency and high-frequency components with the required accuracy. Unfortunately, the only way to obtain information about noise at frequencies that are inaccessible to a particular measuring device is to use multiple devices with different characteristics and then stitch together the resulting PSD. In our experiment, we used a 32-bit $\Delta\Sigma$ ADC for low-frequency (mHz to kHz) measurements at different sampling rates and the Zurich Instruments MFLI Lock-in amplifier~\cite{mfli} for high-frequency (kHz to MHz) measurements. The detailed methodology is described in~\cite{myspie}.

\begin{figure}[h!]
\centering
  \includegraphics[width=0.75\linewidth]{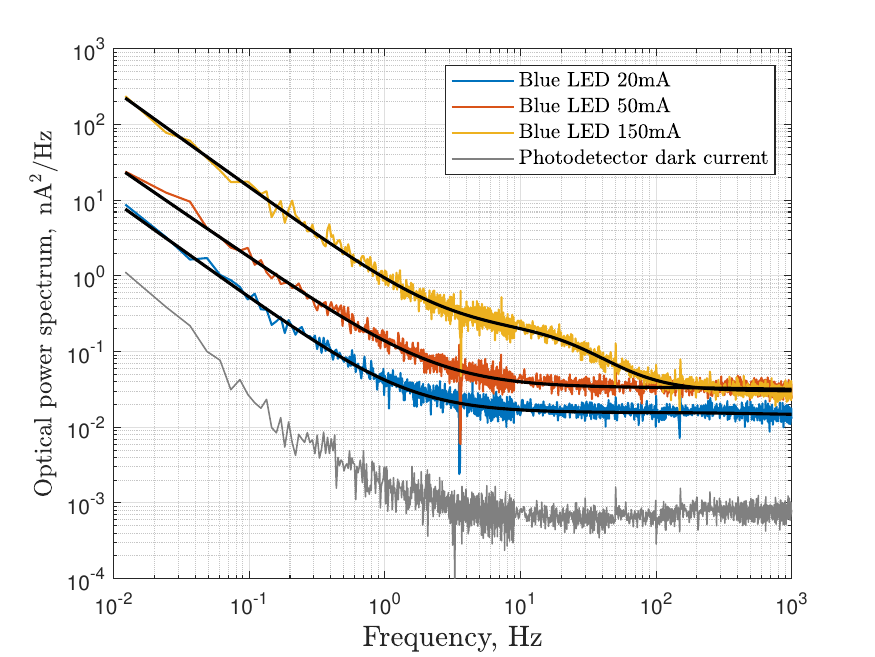}
  \caption{Low-frequency experimental PSD data (blue) for a blue LED with model fit (black) at loglog scale for 20, 50, and 150~mA. The graph also shows the dark current noise of the photodiode (gray), i.e., system noise.}
  \label{fig:optB1}
\end{figure}

The results for currents of 20, 50, and 150~mA are presented on a logarithmic scale in \textbf{Figure~\ref{fig:optB1}}. In addition to the experimental data, the graph shows the system noise, which represents the dark photocurrent of the photodiode (i.e., the sum of photodiode and ADC noise). The experimental data was fitted using {Equation~(\ref{eq:model})}. For currents of 20 and 50~mA, the third part of the equation (Lorentzian) was not considered ($i=0$). For the current of 150~mA, the sum was taken for only one Lorentzian term ($i=1$).

Similar results were obtained in~\cite{ivanov2021,dang2019,ghosh2021}. From {Figure~\ref{fig:optB1}}, it is evident that the low-frequency noise amplitude strongly depends on the current. To approximately calculate the change in flicker noise slope, one can calculate the weighted average of the frequency-dependent exponent $\gamma(f)$. The amplitudes of the first and second terms are used as weights, and their exponents are averaged. In this case, we observe a gradual decrease in the average slope $\langle \gamma(f)\rangle$ as current increases (for the presented currents, $\langle \gamma(f)\rangle_\text{20\,mA}=1.26$ and $\langle \gamma(f)\rangle_\text{50\,mA}=1.14$; this trend continues for intermediate current values). This trend changes when the Lorentzian appears ($\langle \gamma(f)\rangle_\text{150\,mA}=1.29$), which is clearly visible at 150~mA. A change in such a trend indicates that the wide one and narrow one distributions are interrelated. Generally, for all currents $A_{1/f^\gamma}$ is greater than $A_{1/f}$. The gradual emergence of the Lorentzian component from 50 to 150 mA was experimentally confirmed by us. Overall, all trends in the changes of the noise components occur smoothly with the current.

Lorentzian appearance at low frequencies is of particular interest. Similar values of relaxation lifetime ($\tau\approx0.1~\text{s}$) are observed for micrometer-scale regions in the blinking effect~\cite{ruggero2006}. In this case, a discrete change in local point optical intensity is observed between two states at a specific frequency~\cite{shimizu2001}, which is usually in the sub-kilohertz range. The phenomenon is observed in the presence of many defects and increases with increasing carrier concentration. If one of these distributions strongly stands out in the structure, we can observe their influence macroscopically at sufficiently high currents, which could explain the appearance of the Lorentzian at such low frequencies.

Also note that the plateau from the hundreds of hertz to kilohertz shows weak current dependence and does not increase when the Lorentzian appears.

This plateau can be interpreted as white noise or as a plateau of a higher-frequency Lorentzian. In this case, white noise would be the sum of shot noise and thermal noise, the values of which can be theoretically calculated. Using the corresponding formula~\cite{kogan1996} for the current power spectral density $S_\text{shot}=2e\langle I\rangle$ and $S_\text{JN}=4 k_\text{B} T/R$, and converting them into optical fluctuations based on the measured optical flux current dependence, the estimated maximum values are an order of magnitude lower than the system noise $S_\text{opt fund}<10^{-5}~\text{nA}^2/\text{Hz}$ in photocurrent PSD units. This is consistent with expectations~\cite{kogan1996}. The shot noise of the photodiode, caused by the photocurrent flow, is several orders of magnitude lower.

\section{High-frequency noise of blue LED}

\begin{figure}[ht!]
\centering
  \includegraphics[width=\linewidth]{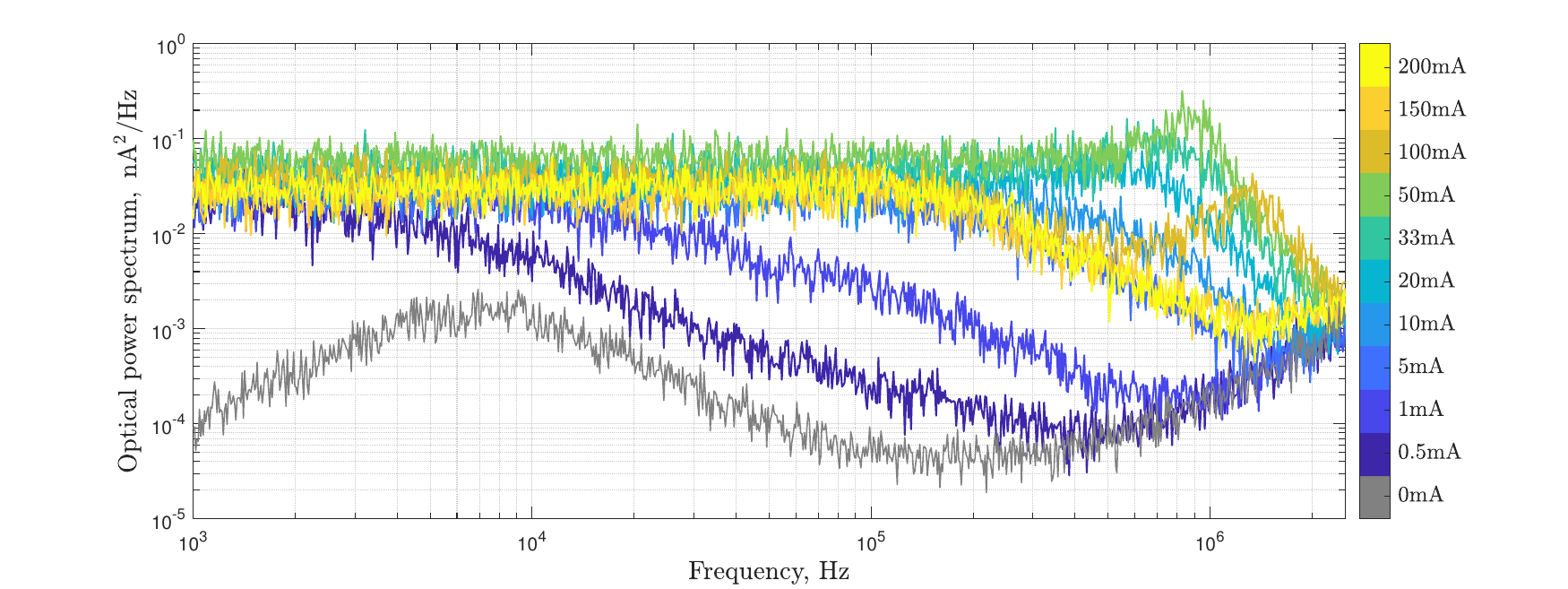}
  \caption{High-frequency experimental PSD data for a blue LED at loglog scale for currents in range from 0.5~mA to 200~mA. The graph also shows the photodiode dark current noise, i.e., 0~mA (gray).}
  \label{fig:highF_B1}
\end{figure}

For the high-frequency part of the same sample shown in {Figure~\ref{fig:optB1}}, a larger series of currents from 0.5~mA to 200~mA was taken due to the more complex nature of the behavior. The results are demonstrated in \textbf{Figure~\ref{fig:highF_B1}}. In addition to the experimental data, the graph shows the photodiode dark current noise. The data was measured using an analog high-pass filter.

In {Figure~\ref{fig:highF_B1}} we see the appearance of a current-dependent Lorentzian at high (around 100kHz) frequencies. So the plateau in the low-frequency range was merely a tail of this high-frequency second Lorentzian. Starting from 0.5~mA, it falls after approximately 2 kHz according to the $f^{-2}$ law up to 0.4 MHz, where system noise starts to dominate. As the frequency increases, the system noise eventually becomes dominant for all curves. It's worth noting that the pronounced peak for the photodiode noise (0 mA) at 0.7 kHz is a peak of the transimpedance amplifier, while the rising peak at 2.5 MHz is the noise of the lock-in amplifier. Note that we do not observe fundamental white noise, as the photodiode noise and LED's own frequency-dependent components prove to be larger. It was verified using neutral density filter that the noise, which is larger than the noise at 0 mA, belongs to the LED, and is not amplified photodiode noise.

The high-frequency Lorentzian strongly depends on current, shifting $\tau$ toward lower values while simultaneously keeping the plateau at the same level. This may indicate either an inverse dependence of amplitude on $\tau$ in {Equation~(\ref{eq:grdot})}, or a special form of $g(\tau)$ that keeps the numerator in {Equation~(\ref{eq:gr})} constant. We can assume that this change in $\tau$ with increasing current indicates that the generation-recombination process now involves capture centers in the active region including barriers that were not involved due to their large distance from the Fermi level before.

Moreover, after 50~mA, a maximum appears at 1 MHz, followed by its gradual shift toward higher frequencies with further current increases. Mathematically, this indicates the presence of a negative Lorentzian, as shown in {Figure~\ref{fig:model}}. The physical explanation could be the so-called ``true recombination''~\cite{rimini1992}, i.e., correlated e-h GR process through traps in the LED, that is described in Appendix~\ref{appendix_gr}. 

Also, after the maximum shifts towards high frequencies over 50~mA, we simultaneously begin to observe the appearance of a low-frequency Lorentzian in Figure~\ref{fig:optB1}. Taking this into account, we can conclude that all these processes are closely interconnected and influence each other.

\section{Noise temperature dependence}

The temperature dependence of the noise can provide information on whether discrete traps or wide continuous distributions of defects contribute to the resulting noise spectrum. 

\begin{figure}[h!]
\begin{center}
  \includegraphics[width=0.75\linewidth]{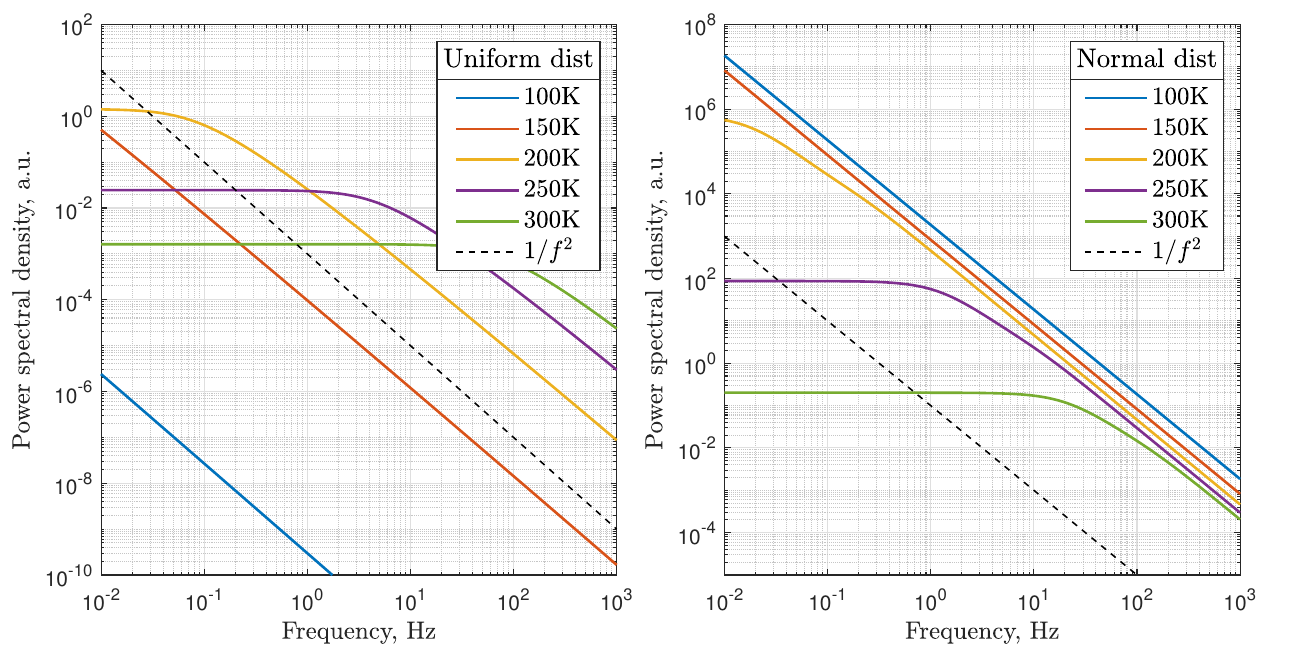}
  \caption{Temperature-dependent noise power spectrum at loglog scale for temperatures 100~K (blue), 150~K (orange), 200~K (yellow), 250~K (purple) and 300~K (green) with $1/f^2$ reference (dashed black). Left: uniform distribution function $g(\tau)$. Right: normal distribution of discrete trap energy depth $g(E)$. The behavior of the curves is typical for the most common values of the energy levels of the traps from~\cite{buffolo2022}: for the left graph, the ranges from the Fermi level were taken as $0.31\dots0.35$ eV, and for the right as $0.33\dots0.37$ eV.}
  \label{fig:tempSim}
  \end{center}
\end{figure}

Considering the trap levels in the band gap as thermally activated and putting the relations between lifetime and corresponding trap energy, we can obtain a temperature dependence for the noise. Its form is defined by the distribution function. Two cases are considered: an uniform relaxation lifetime distribution, similar to the one described in Subsection~\ref{subsection_model}, and a Gaussian distribution of discrete trap energy depth. For details, Appendix~\ref{appendix_tempdep} is provider. Both cases are shown in \textbf{Figure~\ref{fig:tempSim}}.

The proposed models of the temperature dependence of the low-frequency part of the spectrum do not take into account a number of temperature effects, i.e. the temperature dependence of the Fermi level position, carrier mobility (the integral amplitude, which among other things depends on carrier density and mobility, was taken as unity), and, in addition, the dependence of the trap energy distribution (exponential, uniform) in different temperature ranges~\cite{pavesi2006}. These factors will be taken into account in further work.

However, simulated curves in Figure~\ref{fig:tempSim} demonstrate that a decrease in temperature leads to an increase in the amplitude of the resulting superposition of widely distributed GR components. The monotonous change in amplitude with temperature, namely, the absence of a maximum in this dependence confirms that the noise spectrum is not the result of the action of a set of discrete levels, but a wide continuous band in the bandgap~\cite{lukyanchikova1997}.

\begin{figure}[h!]
\centering
  \includegraphics[width=0.9\linewidth]{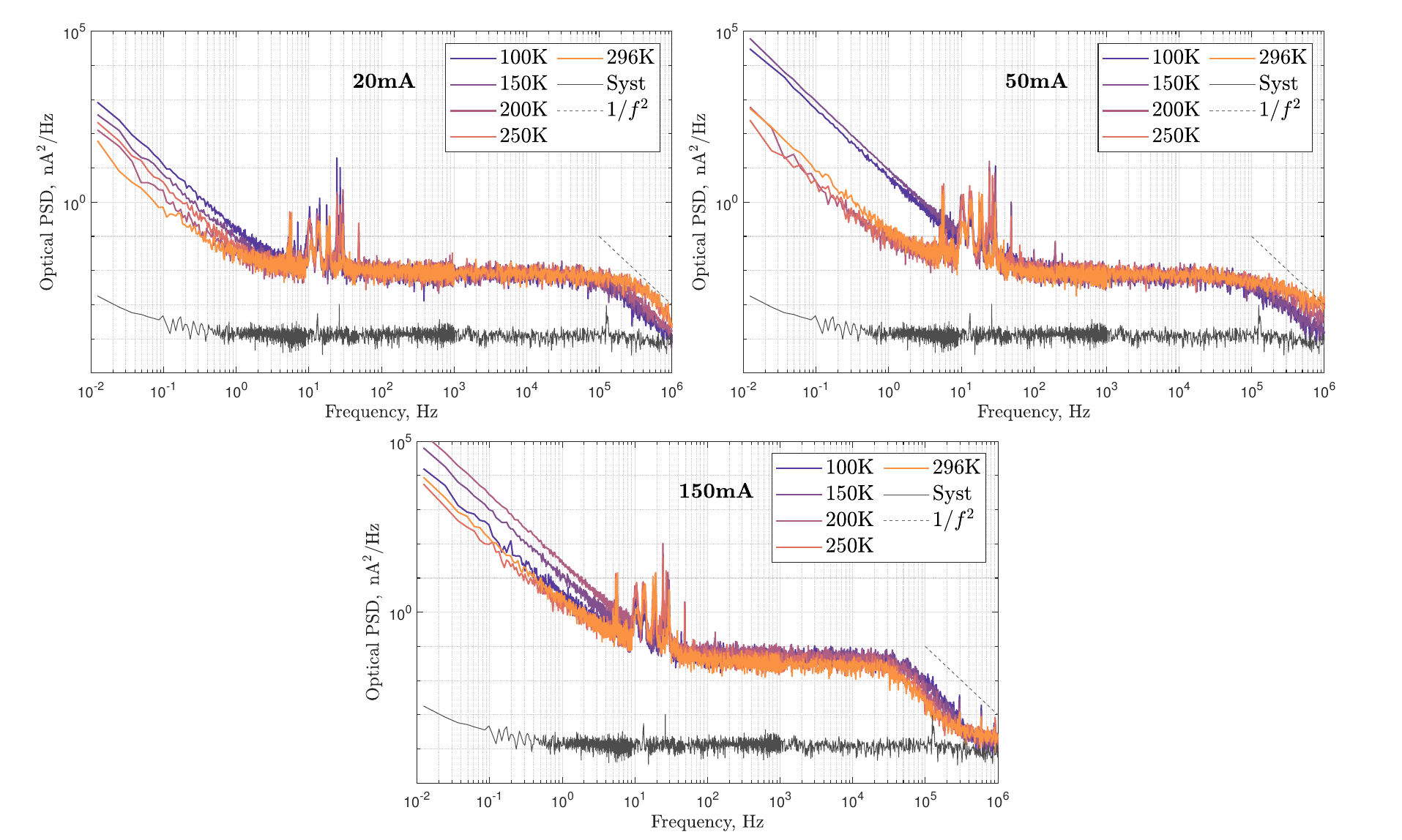}
  \caption{Low- and high-frequency PSD experimental data of a different to the one measured before blue LED at loglog scale for temperatures 100~K (purple), 150~K (violet), 200~K (pink), 250~K (orange) and 296~K (yellow) for currents 20 (top left), 50 (top right) and 150~mA (center bottom). The source of the peaks at 10 Hz is an external factor and they are not related to the intrinsic noise of the LED. The dashed line (gray) marks the $1/f^2$ drop law for comparison purposes.}
  \label{fig:depT_B2}
\end{figure}

The experimental data is presented on \textbf{Figure~\ref{fig:depT_B2}}. It should be noted that this is a different to the one measured before sample but nominally identical than the one shown in Figure~\ref{fig:optB1} and Figure~\ref{fig:highF_B1}.

Firstly, the peaks observed at frequencies around 10 Hz for all currents are systemic noise due to residual mechanical vibration of the cryostat after damping, affecting the optical path. They are not related to real processes in the sample and will be ignored during further analysis. The maximum for system noise at 0.7 kHz is missing, since a different transimpedance amplifier with the same photodiode was used.

Secondly, for low frequencies at 150 mA, we do not see a Lorentzian at room temperature, that was detected in Figure~\ref{fig:optB1}. There is no current dependent Lorentzian shift at high frequencies as in Figure~\ref{fig:highF_B1}. This can indicate a strong difference in defect characteristics between samples even within nominally identical LEDs.

At low currents (20 and 50~mA), the experimental data for low frequencies confirm the simulation prediction with normal distribution of a decrease in the noise amplitude $1/f$ with increasing temperature. 
The coincidence of the slope for up to 1 Hz was observed. At 50~mA, a significant split is observed between 100~K and 150~K and other temperatures. This is explained by the fact that the time needed to reach current equilibrium was longer than the time between the start of measurement and current application, which was 300 s. For other samples, this was sufficient for the LED to reach operating mode, however, during this same time, this drift did not disappear for low temperatures, which gave a large contribution to the $1/f$ noise.

At high current (150~mA), the dependence of the $1/f$ component amplitude on temperature changed and cannot be explained within the framework of the proposed model. The reason of the discrepancy has yet to be determined by further research.

The behavior of the Lorentzian at 0.5 MHz also differs at low and high currents. Firstly, we do not observe a clearly pronounced maximum that was present in Figure~\ref{fig:highF_B1}. Secondly, the slope of the high-frequency Lorentzian changes compared to $1/f^2$, which is shown by the dashed gray line for reference. We see that for high temperatures, the slope significantly differs from $1/f^2$. This may be an intermediate case between wide and narrow distributions in {Equation~(\ref{eq:gr})}, specifically a shift in $\tau_\text{min}$ with a $\tau_\text{max}$, resulting the slope change. For 150~mA, the slope is constant, but there is a small frequency shift when the temperature changes.

\section{Conclusion}

Performed analysis of two commercially available InGaN blue LEDs noise characteristics demonstrates the partial consistency of a semi-empirical model based on the superposition of four distinct spectral noise components. At low frequencies (mHz to kHz), both $1/f^\gamma$ and $1/f$ noise components manifest as a continuous superposition of generation-recombination (GR) processes. The slope and shape of the low-frequency noise spectrum have revealed a dependence on the trap distribution function. The physical explanation should be clarified in further works. This provides a concept for understanding the physical origins of optical noise in LEDs.

In the high-frequency range (kHz to MHz), GR noise contributions from narrow trap intervals, approaching the single-trap case, become dominant. Current dependency studies (20, 50 and 150~mA at low frequencies; from 0.5~mA to 200~mA at high frequencies) reveal that increasing current involves in the GR process capture centers, which are characterized by smaller relaxation lifetimes.

Temperature-dependent measurements (in range from 100~K to 300~K) show the good agreement between experimental data and the suggested model at low current conditions, supporting the proposed idea. However, the high-frequency part of the spectrum shows a deviation from the expected Lorentzian behavior. Furthermore, at high current conditions, notable discrepancies emerge between experimental results and model predictions, suggesting additional noise mechanisms to be considered. Moreover, the fact that we do not observe a Lorentzian at low frequencies for 150 mA for both samples at room temperature, nor a maximum at high frequencies, can indicate a strong difference in defect characteristics between samples even within LEDs which are nominally identical. Three additional nominally identical characterized samples showed differences in their noise behavior.

\appendix
\section{Appendix: Generation-recombination noise}\label{appendix_gr}

\begin{figure}[h!]
\begin{center}
  \includegraphics[width=0.5\linewidth]{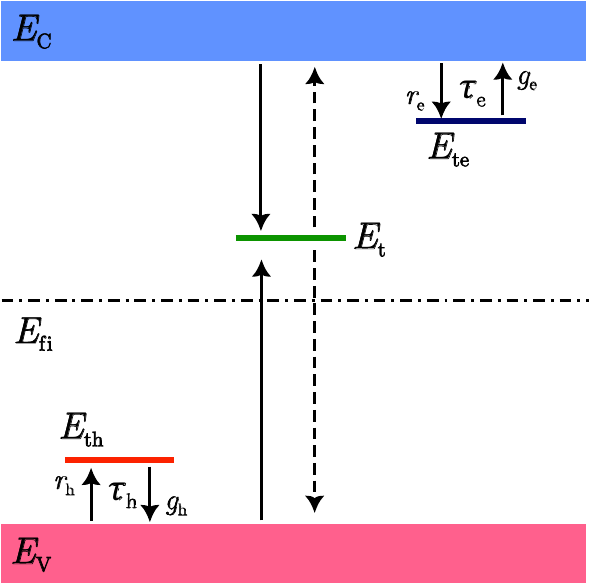}
  \caption{Carrier's transitions. Here $E_\text{fi}$ is the intrinsic Fermi level, $E_\text{C}$ is the conduction band edge, $E_\text{V}$ is the valence band edge, $E_\text{te}$ is the trap level (red) for electrons, $E_\text{th}$ is the trap level (blue) for holes. Generation processes $g_\text{e}$ and $g_\text{h}$ as well as recombination processes $r_\text{e}$ and $r_\text{h}$ are uncorrelated. Green: correlated electron and a hole fluctuations via deep level $E_\text{t}$ of the trap. A captured electron recombines with hole and is not emitted into the conduction band and does not contribute to the current (and current noise) any further. Dashed arrows denote low probability transitions.}
  \label{fig:srh}
  \end{center}
\end{figure}

The random nature of electron transitions between bands and/or local energy levels in the gap causes both fluctuations in the occupancy of the levels and fluctuations in the concentration of free carriers in the bands. If such fluctuations are caused by the statistical nature of the generation and recombination processes, the noise is called generation-recombination (GR) noise. In this case, we refer to generation $(g_\text{e},g_\text{h})$ as the appearance of a free carrier in the corresponding band and to recombination $(r_\text{e},r_\text{h})$ as its disappearance from it, as shown in \textbf{Figure~\ref{fig:srh}}. In the following, we consider only transition to defects but not band-to-band transitions, assuming that their contribution to the resulting fluctuations is rather small~\cite{lukyanchikova1997}.

A number of methods based on different approaches, including statistical and thermodynamic, are used to analyze the fluctuations of charge carriers in a semiconductor~\cite{kogan1996}. Statistical methods are applicable in the case of both equilibrium and non-equilibrium stationary state of the system. The widely used Langevin method is preferred for several advantages. It can be applied to a system that is in a steady state but thermodynamically non-equilibrium. Then if there are two (or more) coupled random variables one has to solve only two (or more) equal to the number of random variables quantity equations for fluctuations. The essence of applying the Langevin method is to add a random force, the so-called Langevin force, to the differential equation describing the kinetics of charge carriers~\cite{vanvliet1971,lukyanchikova1997,kogan1996}. 

In the case of a two-level model, when we consider, for example, only the electron trap $E_\text{te}$ and the conduction band, the Langevin equation is obtained by adding a random noise force $L(t)$ to the continuity equation for electrons~\cite{mitin2002}. Carrier transport is not taken into account in this case. The Fourier spectral transform method can be applied to the Langevin equation, that allows expressing the power spectral density (PSD) of a system response through the spectral density $S_L(f)$ of a random perturbation $L(t)$ of an arbitrary type.

If $N(t)$ is the total number of carriers in the sample and $N_0$ its equilibrium number, we can write the Langevin equation as \textbf{Equation~(\ref{eq:lang})}. Here $\delta N=N-N_0$ is the excessive concentration of carriers fluctuations under the random source $L(t)$ with zero average; $g(N)$ and $r(N)$ are generation and recombination rates respectively, noting that $g$ is decreasing, and $r$ is increasing with $N$.
\begin{equation}\label{eq:lang}
    \dfrac{\text{d}}{\text{d}t}\delta N=g(N)-r(N)+L(t)
\end{equation}

Using series expansion of $g(N)-r(N)$, Equation~(\ref{eq:lang}) yields to \textbf{Equation~(\ref{eq:lang2})}, where $\tau=\left(\frac{\text{d}r}{\text{d}N}-\frac{\text{d}g}{\text{d}N}\right)^{-1}\big|_{N_0}$ at $N_0$.
\begin{equation}\label{eq:lang2}
    \dfrac{\text{d}}{\text{d}t}\delta N=-\left[r'(N_0)-g'(N_0)\right]\delta N+L(T)=-\dfrac{\delta N}{\tau}+L(t)
\end{equation}

Taking the Langevin force as a source of white noise, its correlation function is represented by the product of the delta function and a constant. This constant coefficient at $\delta$ is expressed, according to the Wiener-Khinchin theorem in terms of the spectral density at zero frequency, meaning that $S_L(0)=S_L(f)=4\langle\delta N^2\rangle/\tau=\text{const}$. Thus, we obtain \textbf{Equation~(\ref{eq:lang4})}. 

\begin{equation}\label{eq:lang4}
    S_{\delta N}(f)=\dfrac{4\langle \delta N^2\rangle\tau}{1+(2\pi f)^2\tau^2}
\end{equation}

It is worth noting, that the derivatives of generation $\text{d}g/\text{d}N$ and recombination rates $\text{d}r/\text{d}N$ correspond to the inverse of emission $\tau^{-1}_\text{em}$ and capture times $\tau^{-1}_\text{cap}$ respectively~\cite{mitin2002}. 
This in turn means that $\tau^{-1}=\tau^{-1}_\text{em}+\tau^{-1}_\text{cap}$.

Taken into consideration two types $E_\text{th},E_\text{te}$ of traps (Figure~\ref{fig:srh}) responsible for independent transitions of each type of carriers, the noise spectrum will be a simple superposition of the contributions from electrons and holes and will be written in form of \textbf{Equation~(\ref{eq:lang5})}. Here $\tau_\text{e}^{-1}$ and $\tau_\text{h}^{-1}$ consist of the corresponding electron and holes characteristic relaxation rates.

\begin{equation}\label{eq:lang5}
    S_\text{e-h uncorr}(f)=\dfrac{A_\text{e}\tau_\text{e}}{1+(2\pi f)^2\tau_\text{e}^2}+\dfrac{A_\text{h}\tau_\text{h}}{1+(2\pi f)^2\tau_\text{h}^2}
\end{equation}

The interaction between electrons and holes through recombination centers $E_\text{t}$ i.e., defect energy levels near the middle of the bandgap leads to significant cross-correlation effects, which influence the noise spectra~\cite{rimini1992}. Under non-equilibrium steady state conditions the cross-correlation term between electrons and holes cannot be neglected. When the cross-correlation term is sufficiently negative, it leads to a negative Lorentzian component in the noise spectrum. 

This interaction leads to a reduction in the noise level at certain frequencies, creating a dip or maximum in the noise spectrum. This behavior arises because Shockley-Read-Hall (SRH) recombination at the defect~\cite{srh1952} suppresses the GR noise by extracting charge carrier pairs from the number of fluctuating carriers. This is shown by dashed arrows in Figure~\ref{fig:srh}.

\section{Appendix: Relaxation lifetime distribution}\label{appendix_dist}

Let’s assume that discrete trap is characterized by time constant $\tau_i$. Then discrete multiple-trap levels will be described by \textbf{Equation~(\ref{eq:grsum})}.

\begin{equation}\label{eq:grsum}
    S_{\delta N}(f)=\sum\limits_i\dfrac{4\langle\delta N^2\rangle\tau_i}{(1+\tau_i^2(2\pi f)^2)}
\end{equation}

Assuming that the amplitude of the Lorentzian depends only on the total concentration of charge carriers and their mobility in the semiconductor~\cite{lukyanchikova1997} for the case of a continuous trap distribution, we can write \textbf{Equation~(\ref{eq:grint})}. Here $g(\tau)$ is the probability density function~\cite{vandamme2008}. Since such integral does not converge on the interval $[0;+\infty]$ (the noise energy can not tend to infinity), we will consider it in a limited but arbitrarily wide interval $[\tau_\text{min};~\tau_\text{max}]$. $g(\tau)$ obeys normalization condition $\int_0^\infty g (\tau) \text{d}\tau=1$. 

\begin{equation}\label{eq:grint}
    S_{\delta N}(f)=4 \langle\delta N^2\rangle\int\limits_{\tau_\text{min}}^{\tau_\text{max}}\dfrac{\tau g(\tau)}{1+(2\pi \tau f)^2}\,\text{d}\tau
\end{equation}

The solution of Equation~(\ref{eq:grint}) depends on the distribution function $g(\tau)$. The necessary time constant distribution for $1/f$ noise can originate from the trap activation energy distribution, or result from the tunneling of carriers to trap levels.

One of the first approaches developed by McWhorter~\cite{mcwhorter1957} was to consider tunneling of carriers to trap levels inside surface oxide at diverse depth (distribution in depth). This model is particularly adapted to the conditions in field-effect transistors. Spatially distributed recombination centers in an oxide layer leads to a distribution of time constants, because the tunneling process that determines the occupation of the centers has different time constants depending on the distance to be tunneled through. This leads to $\tau\cdot g(\tau)=\text{const}$. Applying the normalization condition to distribution function and substituting it into the  {Equation~(\ref{eq:grint})}, we will get \textbf{Equation~(\ref{eq:psdexp})}.

\begin{equation}\label{eq:psdexp}
    S_{\delta N}(f)=4\langle \delta N^2\rangle\int\limits_{\tau_\text{min}}^{\tau_\text{max}}\dfrac{1}{1+(2\pi f)^2\tau^2}\,\text{d}\tau=\dfrac{4\langle \delta N^2\rangle}{2\pi f\log(\tau_\text{max}/\tau_\text{min})}\left(\arctan(2\pi f\tau_\text{max})-(\arctan2\pi f\tau_\text{min})\right)
\end{equation}

Such a distribution could be considered as a surface contribution to noise. PSD will be proportional to $1/f$ in case of $\tau_\text{max}^{-1}< f<\tau_\text{min}^{-1}$, constant if $f\ll\tau_\text{min}^{-1}$ and proportional to $1/f^2$ if $f\ll\tau_\text{max}^{-1}$. For LEDs, the parameter $\gamma$ for the dependence $S(f)\propto1/f^\gamma$ can be different from 1 or 2. Models like the model of Kirton and Uren~\cite{kirtonuren1989}, which are extensions of the McWhorter model and work well for field-effect transistors, still give $\gamma\approx1$~\cite{vandamme2008} and therefore cannot describe the main part of noise in LEDs.

Instead of using $g(\tau)\propto 1/\tau$, we propose the assumption that the form of $g(\tau)$ provides not discrete values of $\gamma$ but a continuous dependence $\gamma(f)$. One of the approaches may be an uniform distribution of characteristic relaxation lifetimes constants, as written in \textbf{Equation~(\ref{eq:uniform})} considering the normalization condition.
\begin{equation}\label{eq:uniform}
    \begin{cases}
    g(\tau)\,\text{d}\tau=\text{const}\cdot\,\text{d}\tau=\dfrac{\text{d}\tau}{\tau_\text{max}-\tau_\text{min}}=\dfrac{\text{d}\tau}{\Delta\tau} &\text{for}\ \tau_\text{min}<\tau<\tau_\text{max}\\
    g(\tau)\,\text{d}\tau=0 &\text{otherwise}
    \end{cases}
\end{equation}

Substituting Equation~(\ref{eq:uniform}) in Equation~(\ref{eq:grint}), we can write \textbf{Equation~(\ref{eq:uniform2})}.
\begin{align}\label{eq:uniform2}
\begin{split}
    S_{\delta N}(f)=&4 \langle \delta N^2\rangle\int\limits_{\tau_\text{min}}^{\tau_\text{max}}\dfrac{\tau (\Delta\tau)^{-1}}{1+(2\pi \tau f)^2}\,\text{d}\tau=\dfrac{2\langle \delta N^2\rangle}{(2\pi f)^2\Delta\tau}\left(\log(1+(2\pi f\tau_\text{max})^2)-\log(1+(2\pi f\tau_\text{min})^2)\right)=\\&=
    \begin{cases}
   \propto 1/f^{\gamma(f)} &\text{for}\ \tau_\text{max}^{-1}< f<\tau_\text{min}^{-1}\\
    \propto{\tau_\text{max}^2}/{\Delta\tau} &\text{for}\ f\ll\tau_\text{max}^{-1}\\
    \propto 1/f^{2} &\text{for}\ f\gg\tau_\text{min}^{-1}\\
    \end{cases}
\end{split}
\end{align}

\begin{figure}[h!]
\begin{center}
  \includegraphics[width=0.75\linewidth]{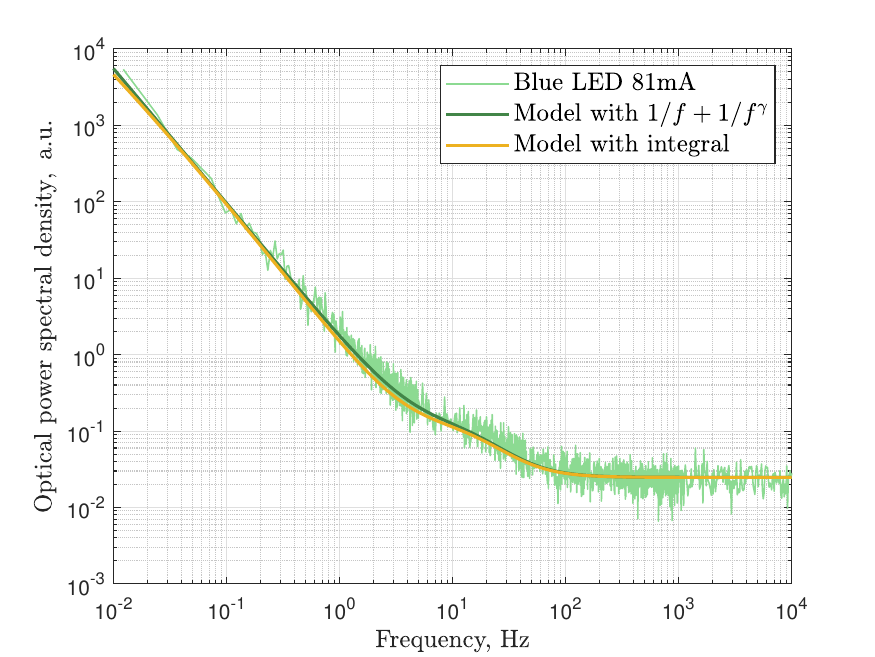}
  \caption{Fitting of the experimental data by the empirical model and the proposed integral for a blue LED with a current of 81 mA. Light green line denotes experimental data, dark green denotes the model that corresponds to Equation~(\ref{eq:model}), and yellow denotes the same model, but instead of the empirical $1/f$ and $1/f^\gamma$ components, a single integral with the form of Equation~(\ref{eq:uniform2}) with $\tau_\text{max}=10^3$~s and $\tau_\text{min}=10^{-1}$~s was taken. For $10^{-2}~\text{Hz}\leq f \leq 3$ Hz, the slope is equal to $\gamma\approx1.7$.}
  \label{fig:fittingInt}
  \end{center}
\end{figure}

This result looks like the Lorentzian, except with a more complicated frequency dependence that does not immediately fall off after the characteristic frequency as $1/f^2$. In this case frequency dependent $\gamma$ part strongly depends on the $\tau_\text{min}$ and its average $\langle\gamma(f)\rangle$ increases as $\tau_\text{min}$ increases. Depending on $\tau_\text{max}$ and $\tau_\text{min}$, $\gamma$ can take values between 1.5 and 2. The amplitude is mostly defined by $\tau_\text{max}$. The characteristic frequency is determined by $\tau_\text{max}$: as it increases, the characteristic frequency decreases. Thus, an integral can describe the $1/f^\gamma$ part if its plateau is at very low frequencies $(\tau_\text{max}~\text{is large})$ as it is shown in \textbf{Figure~\ref{fig:fittingInt}}. If the difference between $\tau_\text{max}$ and $\tau_\text{min}$ is small $\tau_\text{max}=\tau_\text{min}(1+\epsilon),~\epsilon\to0$, the integral in Equation~(\ref{eq:grint}) looks like the discrete Lorentzian in the form of Equation~(\ref{eq:lang2}).

\section{Appendix: Noise temperature dependence model}\label{appendix_tempdep}

The behavior of the spectral model as a function of temperature is of interest. We will observe the change in the noise spectrum from uniformly distributed traps in the low-frequency region. In addition, for comparison, we will use the normal energy distribution for one discrete trap. 



Let’s describe the time constant dependence on temperature. It was earlier mentioned that relaxation time for low frequencies can be equal to SRH lifetime. Without loss of generality, in our case we used the simplest conventional SRH lifetime formula in form of \textbf{Equation~(\ref{eq:srh})}~\cite{buffolo2022} to explore the temperature effect in the simplest way under the assumptions of equality of holes and electrons capture rates, and small deviations of carrier densities from equilibrium within an intrinsic material. Here $T$ is the absolute temperature, $E_\text{t}$ is the trap energy depth, $E_\text{fi}$ is the intrinsic Fermi level and $1/\tau_\text{0n,p}=N_\text{t}v_\text{tn,p}\sigma_\text{n,p}$, where $v_\text{tn,p}$ is the electrons and holes thermal velocity, $\sigma_\text{n,p}$ is the respective apparent carrier capture
cross section of the trap and $N_\text{t}$ is the traps density.
\begin{equation}\label{eq:srh}
    \tau_\text{SRH}=\tau_\text{0n,p}\left[1+\cosh\left(\dfrac{E_\text{t}-E_\text{fi}}{k_\text{B}T}\right)\right]
\end{equation}
We used typical parameters for InGaN LED defects and impurities  from~\cite{buffolo2022,casu2022}.
Assuming that distribution has the form of Equation~(\ref{eq:uniform}), we can write \textbf{Equation~(\ref{eq:dtaude})}
\begin{equation}\label{eq:dtaude}
    g(E)\,\text{d}E=g(\tau)\,\text{d}\tau=g(\tau)\dfrac{\text{d}\tau}{\text{d}E}\,\text{d}E=\dfrac{1}{\tau_\text{max}-\tau_\text{min}}\cdot \dfrac{\tau_{0}}{(k_\text{B}T)}\sinh\left( \dfrac{E}{k_\text{B}T}\right)\,\text{d}E
\end{equation}

Substituting Equation~(\ref{eq:dtaude}) into Equation~(\ref{eq:grint}) and replacing $\tau$ according to Equation~(\ref{eq:srh}), we get  \textbf{Equation~(\ref{eq:tmodelApp})}, where $E$ means $E_\text{t}-E_\text{fi}$. Here $\alpha_{1,2}=\cosh(E_{1,2}/k_BT)$, $\beta=\tau_\text{0n,p}\cdot\sqrt{T_\text{room}}$, $k_\text{B}$ is the Boltzmann constant, $E_{1,2}$ are the energy $E$ distribution boundaries and $A$ includes weak temperature and energy dependent parameters.
\begin{equation}\label{eq:tmodelApp}
	S_\text{GR}(f,T)=A\cdot\int\limits_{E_1}^{E_2}\frac{\beta}{2\sqrt{T}}\left(\frac{2\sinh(E/k_\text{B}T)+\cosh(2E/k_\text{B}T)/2}{(\alpha_2-\alpha_1)\left(1+(2\pi f \beta)^2/T\cdot(1+\cosh(E/k_\text{B}T)^2)\right)}\right)\,\text{d}E
\end{equation}

In case of single trap, instead of taking uniform distribution, we should use normal distribution in form of \textbf{Equation~(\ref{eq:normalDist})}. Here $\sigma=(E_2-E_1)/6$ is the variance and $E_0=(E_2+E_1)/2$ the mean trap position. 
\begin{equation}\label{eq:normalDist}
	g(E)=\dfrac{1}{\sqrt{2\pi\sigma^2}}\exp\left[-\dfrac{(E-E_0)^2}{2\sigma^2}\right]
\end{equation}

Similar to Equation~(\ref{eq:tmodelApp}), using Equation~(\ref{eq:srh}) and substituting all into Equation~(\ref{eq:grint}), we obtain Equation~(\ref{eq:tmodelNorm}).
\begin{equation}\label{eq:tmodelNorm}
	S_\text{GR}(f,T)=A\cdot\int\limits_{E_1}^{E_2}\left(\frac{(1+\cosh(E/k_\text{B}T))\exp\left[-\frac{(E-E_0)^2}{2\sigma^2}\right]}{\sqrt{2\pi\sigma^2}\left(1+(2\pi f )^2(1+\cosh(E/k_\text{B}T))^2\right)}\right)\,\text{d}E
\end{equation}





%
\bibliographystyle{MSP}
\bibliography{MSP-bib-v2.1.bib}

\begin{thebibliography}{10}
\providecommand{\url}[1]{\texttt{#1}}
\providecommand{\urlprefix}{URL }

\bibitem{koike2002}
M.~Koike, N.~Shibata, H.~Kato, Y.~Takahashi,
\newblock \emph{IEEE Journal of Selected Topics in Quantum Electronics}
  \textbf{2002}, \emph{8}, 2 271.

\bibitem{glemza2018}
J.~Glemza, J.~Matukas, S.~Pralgauskaite, V.~Palenskis,
\newblock \emph{Lithuanian Journal of Physics} \textbf{2018}, \emph{58} 194.

\bibitem{bychikhin2005}
S.~Bychikhin, D.~Pogany, L.~K.~J. Vandamme, G.~Meneghesso, E.~Zanoni,
\newblock \emph{Journal of Applied Physics} \textbf{2005}, \emph{97}, 12
  123714.

\bibitem{dobrzanski2004}
L.~Dobrzanski,
\newblock \emph{Journal of Applied Physics} \textbf{2004}, \emph{96}, 8 4135.

\bibitem{palenskis2021}
V.~Palenskis, J.~Matukas, J.~Glemza, S.~Pralgauskaite,
\newblock \emph{Materials (Basel)} \textbf{2021}, \emph{15}, 1 13.

\bibitem{pralgauskaite2015}
S.~Pralgauskaite, V.~Palenskis, J.~Matukas, J.~Glemza, G.~Muliuk, B.~Saulys,
  A.~Trinkunas,
\newblock \emph{Microelectronics Reliability} \textbf{2015}, \emph{55}, 1 52.

\bibitem{rumyantsev2005}
S.~L. Rumyantsev, S.~Sawyer, N.~Pala, M.~S. Shur, Y.~Bilenko, J.~P. Zhang,
  X.~Hu, A.~Lunev, J.~Deng, R.~Gaska,
\newblock In A.~A. Balandin, F.~Danneville, M.~J. Deen, D.~M. Fleetwood,
  editors, \emph{Noise in Devices and Circuits III}, volume 5844. International
  Society for Optics and Photonics, SPIE, \textbf{2005} 75 -- 85.

\bibitem{palenskis2011}
V.~Palenskis, B.~Saulys, J.~Matukas, S.~Pralgauskaitė, G.~Kulikauskas,
\newblock \emph{Acta Physica Polonica Series A} \textbf{2011}, \emph{119} 514.

\bibitem{sawyer2006}
S.~Sawyer, S.~L. Rumyantsev, M.~S. Shur, N.~Pala, Y.~Bilenko, J.~P. Zhang,
  X.~Hu, A.~Lunev, J.~Deng, R.~Gaska,
\newblock \emph{Journal of Applied Physics} \textbf{2006}, \emph{100}, 3
  034504.

\bibitem{vanderziel1979}
A.~{Van Der Ziel},
\newblock \emph{Flicker Noise in Electronic Devices}, volume~49 of
  \emph{Advances in Electronics and Electron Physics},
\newblock Academic Press, New York, NY, \textbf{1979}.

\bibitem{kogan1996}
S.~Kogan,
\newblock \emph{Generation–recombination noise}, 118–129,
\newblock Cambridge University Press, \textbf{1996}.

\bibitem{machlup1954}
S.~Machlup,
\newblock \emph{Journal of Applied Physics} \textbf{1954}, \emph{25}, 3 341.

\bibitem{vandamme2008}
L.~K.~J. Vandamme, F.~N. Hooge,
\newblock \emph{IEEE Transactions on Electron Devices} \textbf{2008},
  \emph{55}, 11 3070.

\bibitem{rimini1992}
M.~Rimini-D{\"o}ring, A.~Hangleiter, S.~Winkler, N.~Kl{\"o}tzer,
\newblock \emph{Applied Physics A} \textbf{1992}, \emph{54} 120.

\bibitem{lukyanchikova1997}
N.~Lukyanchikova, B.~Jones,
\newblock \emph{Noise Research in Semiconductor Physics},
\newblock Taylor \& Francis, \textbf{1997}.

\bibitem{mitin2002}
V.~Mitin, L.~Reggiani, L.~Varani,
\newblock In A.~Bajandin, editor, \emph{Noise and Fluctuations Control in
  Electronic Devices}. American Scientific Publishers, \textbf{2002} .

\bibitem{vanderziel1988}
A.~van~der Ziel,
\newblock \emph{Proceedings of the IEEE} \textbf{1988}, \emph{76}, 3 233.

\bibitem{mcwhorter1957}
A.~McWhorter,
\newblock \emph{Semiconductor surface physics} \textbf{1957}, 207--228.

\bibitem{kirtonuren1989}
M.~J. Kirton, M.~J. Uren,
\newblock \emph{Advances in Physics} \textbf{1989}, \emph{38}, 4 367.

\bibitem{bisquert2008}
J.~Bisquert,
\newblock \emph{Phys. Rev. B} \textbf{2008}, \emph{77} 235203.

\bibitem{mfli}
Zurich Instruments AG, Zurich,
\newblock \emph{MFLI User Manual 500 kHz / 5 MHz Lock-in Amplifier}, 25.01
  edition, \textbf{2025}.

\bibitem{myspie}
D.~Bohomolov, V.~Ivanova, U.~T. Schwarz,
\newblock In \emph{Gallium Nitride Materials and Devices XIX}, volume 12886.
  SPIE, \textbf{2024} 67--75.

\bibitem{ivanov2021}
A.~M. Ivanov,
\newblock \emph{Technical Physics} \textbf{2021}, \emph{66}, 1 71.

\bibitem{dang2019}
W.~Dang-Hui, X.~Tian-Han,
\newblock \emph{ACTA PHYSICA SINICA} \textbf{2019}, \emph{68}, 12.

\bibitem{ghosh2021}
S.~Ghosh, K.~Fu, F.~Kargar, S.~Rumyantsev, Y.~Zhao, A.~A. Balandin,
\newblock \emph{Applied Physics Letters} \textbf{2021}, \emph{119}, 24 243505.

\bibitem{ruggero2006}
R.~Micheletto, M.~Abiko, A.~Kaneta, Y.~Kawakami, Y.~Narukawa, T.~Mukai,
\newblock \emph{Applied Physics Letters} \textbf{2006}, \emph{88}, 6, 061118.

\bibitem{shimizu2001}
K.~T. Shimizu, R.~G. Neuhauser, C.~A. Leatherdale, S.~A. Empedocles, W.~K. Woo,
  M.~G. Bawendi,
\newblock \emph{Phys. Rev. B} \textbf{2001}, \emph{63} 205316.

\bibitem{buffolo2022}
M.~Buffolo, A.~Caria, F.~Piva, N.~Roccato, C.~Casu, C.~De~Santi, N.~Trivellin,
  G.~Meneghesso, E.~Zanoni, M.~Meneghini,
\newblock \emph{physica status solidi (a)} \textbf{2022}, \emph{219}, 8
  2100727.

\bibitem{pavesi2006}
M.~Pavesi, M.~Manfredi, F.~Rossi, M.~Meneghini, E.~Zanoni, U.~Zehnder,
  U.~Strauss,
\newblock \emph{Applied Physics Letters} \textbf{2006}, \emph{89}, 4 041917.

\bibitem{vanvliet1971}
K.~M. van Vliet,
\newblock \emph{Journal of Mathematical Physics} \textbf{1971}, \emph{12}, 9
  1998.

\bibitem{srh1952}
W.~Shockley, W.~T. Read,
\newblock \emph{Phys. Rev.} \textbf{1952}, \emph{87} 835.

\bibitem{casu2022}
C.~Casu, M.~Buffolo, A.~Caria, C.~De~Santi, E.~Zanoni, G.~Meneghesso,
  M.~Meneghini,
\newblock \emph{Micromachines} \textbf{2022}, \emph{13}, 8.

\end{thebibliography}

\end{document}